# Kinetic Freeze-out Spectra of Identified Particles Produced in p-Pb Collisions at $\sqrt{s_{NN}}$ = 5.02 TeV

Inam – ul Bashir[1*] and Saeed Uddin[1]

[1]Department of Physics, Jamia Millia Islamia, New Delhi – 110025, INDIA

* email: inamhep@gmail.com

## Abstract

We study the transverse momentum spectra of identified pions ($\pi^- + \pi^+$), Kaons (($K^- + K^+$), $K_0^s$), protons ($p + \bar{p}$) and Lambdas ($\Lambda + \bar{\Lambda}$) produced at mid-rapidity (0 < y < 0.5) in most central p-Pb collisions at $\sqrt{s_{NN}}$ = 5.02 TeV in comparison with a Unified Statistical Thermal Freeze-out Model (USTFM). The measurements of pions are reported upto $p_T$ = 3 GeV, the kaons ($K^- + K^+$) are reported upto $p_T$ = 2.5 GeV, $K_0^s$ is reported upto $p_T$ = 7 GeV and the baryons (protons and Lambdas) are reported upto $p_T$ = 3.5 GeV. A good agreement is seen between the calculated results and the experimental data points taken from the ALICE experiment. The transverse momentum spectra are found to be flatter for heavy particles than for light particles. Bulk freeze-out properties in terms of kinetic freeze-out temperature and the transverse collective flow velocity are extracted from the fits of the transverse momentum spectra of these hadrons. The effect of resonance decay contributions has also been taken care of.

## Introduction

The statistical and hydrodynamic models predicted the existence of a hot and dense matter produced in the initial stage, which rapidly expands and cools down, ultimately undergoing a transition to a hadron gas phase [1]. A collective hydrodynamic flow developed from the initially generated pressure gradients results in a characteristic dependence of the shape of the transverse momentum ($p_T$) distribution on the particle mass, which can be described with a kinetic freeze-out temperature T and a collective expansion velocity $\beta_T$ [2]. The interpretation of the heavy-ion results depends crucially on the comparison with results from smaller collision systems such as proton-proton (p-p) or proton-nucleus (p-A). Proton-nucleus (p-A) collisions are intermediate between proton-proton (p-p) and nucleus-nucleus (A-A) collisions in terms of the system size and the number of produced particles. Particle production in p-A collisions, in contrast to p-p, is expected to be sensitive to nuclear effects in the initial state. Comparing particle production in p-p, p-A and A-A reactions has frequently been used to separate the initial state effects from the final state effects. Measurements in p-Pb collisions at LHC allow one to attribute the final state effects to the formation of hot QCD matter in heavy ion collisions [3]. The $p_T$ distributions and yields of particles of different mass at low and intermediate momenta of less than a few GeV (~ 3 - 4 GeV) can provide important information about the system created in high energy hadron reactions. This is because a vast majority of the particles is produced in this soft region where a thermally equilibrated system is expected to be formed. The measurement of charged kaons is a significant tool to further understand the thermalization of the system and the mechanism of strangeness production in these collisions. In our previous analysis [4, 5], we have studied the particle production in p-p and Pb-Pb collisions at LHC by employing the phenomenological USTFM approach. Significant collective flow effects were seen in both the cases which gave a support to the assumption of complete thermalization of the produced system in these collisions at LHC. It will therefore be interesting to study the medium properties of the system produced in p-Pb collisions which can be treated as the intermediate between p-p and p-Pb collisions. In order to address the particle production in the QCD matter produced in p-Pb collisions, a systematic study of the identified particles over a broad $p_T$ range is required. We, therefore in this analysis, have used the same phenomenological approach to reproduce the mid-rapidity (0<y<0.5) $p_T$-distributions of identified particles produced in p-Pb collisions at the LHC energy of $\sqrt{s_{NN}}$ = 5.02 TeV. The model [4 - 9] incorporates the effects of both longitudinal as well as transverse hydrodynamic flow in the produced system. We have taken care of the resonance decay contributions to a given hadronic specie.

## Results and Discussion

The details of the model can be found from the references [4 – 9]. Assuming that the system attains a thermo-chemical equilibrium at freeze-out, the momentum distributions of hadrons, emitted from within an expanding fireball, are characterized by the Lorentz-invariant Cooper-Frye formula [10]

$$E\frac{d^3n}{d^3P} = \frac{g}{(2\pi)^3}\int f\left(\frac{p^\mu u^\mu}{T},\lambda\right) p^\mu d\Sigma_\mu, \quad (1)$$

Where $\Sigma_f$ represents a 3-dimensional freeze-out hyper-surface and 'g = 2J+1' is the degree of degeneracy of the expanding relativistic hadronic gas. After incorporating the effects of resonance decay contributions, the final particle transverse momentum spectral shape takes the form

$$E'\frac{d^3N}{d^3p'} =$$

$$\frac{1}{2p'}\left\{\frac{m_h}{p^*}\right\}\lambda_h\, g_h\, e^{-\alpha\theta E'E^*}\left\{\frac{\alpha}{\theta}\left[E'E^*\,Sinh(\alpha\theta p'p^*) - p'p^*\,Cosh(\alpha\theta p'p^*)\right] + T^2 Sinh(\alpha\theta p'p^*)\right\}, \quad (2)$$

where $\alpha$ and $\theta$ are given by $m_h/m^2$ and $1/T$, respectively. The invariant cross section will have the same value in all the Lorentz frames, i.e

$$E\frac{d^3N}{d^3p} = E'\frac{d^3N}{d^3p'} \quad (3)$$

The subscript $h$ in equation 2 stands for the decaying (parent) hadron. The two body decay kinematics gives the *product* hadron's momentum and energy in the "rest frame of the *decaying hadron*" as $p^* = (E^{*2} - m^2)^{1/2}$ and $E^* = \frac{m_h^2 - m_j^2 + m^2}{2m_h}$ where $m_j$ indicates the mass of the *other* decay hadron produced along with the first one. The *transverse* velocity component of the hadronic fireball, $\beta_T$ is assumed to vary with the transverse coordinate $r$ in accordance with the Blast Wave model as $\beta_T(r) = \beta_T^s\left(\frac{r}{R}\right)^n$ where $n$ is an index which fixes the profile of $\beta_T(r)$ in the transverse direction and $\beta_T^s$ is the hadronic fluid *surface transverse expansion velocity* and is fixed in the model by using the parameterization $\beta_T^s = \beta_T^0\sqrt{1-\beta_z^2}$. This relation is also required to ensure that the net velocity $\beta$ of any fluid element must satisfy $\beta = \sqrt{\beta_T^2 + \beta_z^2} < 1$.

The $p_T$ distributions of pions ($\pi^-$ + $\pi^+$), Kaons (($K^-$ + $K^+$), $K_0^s$), protons (p + $\bar{p}$) and Lambdas ($\Lambda$ + $\bar{\Lambda}$) produced at mid-rapidity (0 < y < 0.5) in most central p-Pb collisions at $\sqrt{s_{NN}}$ = 5.02 TeV are shown below in Figure 1. In our analysis, we have assumed the mid-rapidity baryon chemical potential to be ~ 0, [5, 11] under the assumption of a baryon symmetric matter expected to be formed under the condition of a high degree of nuclear transparency in the nucleus-nucleus collisions at LHC energy, i.e Bjorken's approach. We employ the minimum $\chi^2$/dof method to fit the experimental data taken from ALICE experiment [12]. The $p_T$- distributions are found to be *insensitive* to the value of $\sigma$ (the width of the matter distribution) thus it is set equal to 5 in our model [5]. As shown in Figure 1, a good agreement is seen between the experimental data points (shown by grey filled squares) and the model predictions (shown by black solid curves) suggesting the statistical nature of the particle production and hence the validity of our approach. However a slight disagreement is seen between the theoretical results and the experimental data at higher values of transverse momenta $p_T$. It is because that the statistical hydrodynamic calculations cannot describe the hadron spectra at such large transverse momenta. The hadrons detected in this region are essentially formed by the partons which are a result of the hard processes. These originate from the direct fragmentation of high-energy partons of the colliding beams and therefore are not able to thermalize through the process of multiple collisions [5]. The various freeze-out conditions obtained by fitting the $p_T$-distributions of these hadrons are given in Table 1 below. It is seen from the table that a significant value of collective flow is observed for all the studied particles in case of p-Pb collisions at LHC. This hints towards the thermalization of the produced system in p-Pb collisions and the possible formation of Quark Gluon Plasma in p-Pb collisions at LHC [13]. The possibility of collective flow effects in p-Pb collisions was also suggested by the results reported in the references [14 - 17]. The collective flow velocity decreases and the thermal freeze-out temperature increases on going from lighter mass particles to heavier particles. This trend, known as sequential freeze-out, is similar to that found in case of Pb-Pb collisions at LHC [5] and to that found in Au-Au collisions at RHIC [7]. However it is not so in case of p-p collisions at LHC [4] where the phenomena of sequential freeze-out is not much apparent. The reason for this sequential/systematic freeze-out of the hadrons in p-Pb collisions can be attributed to an early freeze-out for the massive particles (hyperons) when the thermal temperature is high and the collective flow is in the early stage of development and consequently $\beta_T^0$ is small. The early freeze-out of these particles is due to their smaller cross-section with the hadronic matter. Also, the similar freeze-out conditions of kaons indicate their near simultaneous freeze-out from the system. A comparison of the freeze-out parameters with those obtained at RHIC [6, 7] indicates that the collective flow increases considerably where as the freeze-out temperature decreases significantly on going from RHIC to LHC energies. This fact is understood to be due to the large energies available for particle production at LHC which results in a larger collective flow and smaller freeze-out temperatures. While the spectra are mainly determined by thermal freeze-out temperature T and the transverse flow velocity $\beta_T^0$, the shape of the flow velocity profile also has some effect on the spectra, due to the non-linearity in the dependence of the spectral shape on the flow velocity. This is indicated by the different values of the velocity profile index $n$ for the different particle spectra. The value of transverse flow velocity parameter $n$ is found to decrease from lower mass particles to higher mass particles. This trend was also observed in Pb-Pb collisions at LHC [5]. The transverse momentum spectra exhibit a broadening for heavier particles, which has also been observed in Pb-Pb collisions at LHC [5] and in Au-Au collisions at RHIC [7]. This broadening of the $p_T$ spectra find their natural explanation in the collective expansion of the system [18]. The resemblance of the sequential freeze-out scenario and the $p_T$ spectra broadening of p-Pb collision system with those of Pb-Pb and Au-Au collision systems at LHC and RHIC make us believe that the p-Pb collision system behaves more like a heavy ion system rather than a system produced in hadron-hadron interactions. This, in addition with the collective flow signatures, provides the support for the validity of our approach to p-A collision systems at LHC.

| Particle | $\beta_T^0$ | T (MeV) | n |
|---|---|---|---|
| Pion | 0.95 | 75.0 | 2.74 |
| Proton | 0.83 | 98.0 | 1.33 |
| $K_0^S$ | 0.80 | 132.0 | 2.33 |
| Kaon | 0.81 | 131.0 | 1.50 |
| Lambda | 0.78 | 160.0 | 1.12 |

**Table 1**. *Kinetic freeze-out parameters of hadrons*

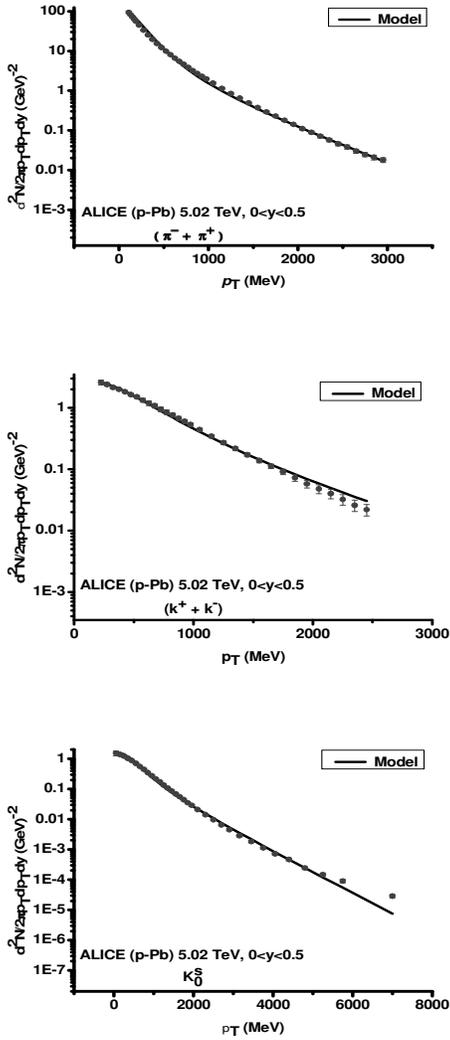

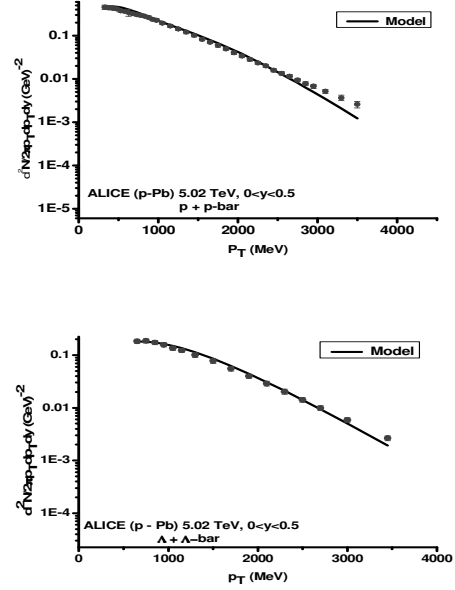

**Figure. 1** $p_T$ – *distributons of various hadrons produced in p-Pb collisions at $\sqrt{s_{NN}}$ = 5.02 TeV. Grey filled circles represent data points and black solid line represent theoretical curve. Errors are the sum of systematic and statistical uncertainties.*

## Conclusion

In summary, we have successfully reproduced the mid-rapidity $p_T$ spectra of identified pions ($\pi^- + \pi^+$), Kaons (($K^- + K^+$), $K_0^s$), protons ($p + \bar{p}$) and Lambdas ($\Lambda + \bar{\Lambda}$) produced in central p-Pb collisions at LHC in comparison of Unified Statistical Thermal Freeze-out Model (USTFM). A good agreement between the theoretical results and the experimental data hints at the complete thermalization of the produced system and hence the validity of our approach. The thermal freeze-out conditions show the phenomena of a sequential freeze-out of the various particles as seen in heavy ion collisions at RHIC and LHC. Significant collective flow is found in the system produced in p-Pb collisions and this hints at the possible formation of the QGP in these collisions. Kaons are found to freeze-out simultaneously from the system.

## Acknowledgement

The authors are grateful to the University Grants Commission for the financial assistance.

## References

[1] B. Muller and J. L. Nagle, Ann. Rev. Nucl. Part. Sci. 56, 93 (2006).

[2] E. Schnedermann, J. Sollfrank, and U. W. Heinz, Phys. Rev. C48, 2462 (1993).

[3] C. Salgado, J. Alvarez-Muniz, F. Arleo, N. Armesto, M. Botje, et al., J.Phys.G G39 (2012) 015010

[4] Inam-ul Bashir, Riyaz Ahmed Bhat and Saeed Uddin, arXiv:1510.05894 [hep-ph]

[5] S. Uddin et al., Advances in High energy physics, 2015, Article ID 154853, 7 pages (2015)

[6] S. Uddin et al., J. Phys. G 39 015012 (2012)

[7] S. Uddin et al., Nuclear Physics A 934 121–132 (2015)

[8] Inam-ul Bashir et al., Int. Journal of Mod. Phys. A, Vol. 30 (2015) 1550139 (11 pages)

[9] Inam-ul Bashir et al., Journal of Experimental and Theoretical Physics, 2015, Vol. 121, No. 2, pp. 206–211

[10] F. Cooper and G. Frye, 1974, *Phys. Rev. D,* 186-189

[11] J. D. Bjorken, Phys. Rev. D27, 140 (1983).

[12] B. Abelev et. al., ALICE collaboration, Phys. Lett. B 728 (2014) 25-38

[13] S.A. Bass et al., Nucl. Phys. A661, 205 (1999).

[14] A. Adare et al., (PHENIX Collaboration, arXiv:1303.1794 (2013).

[15] S. Chatrchyan et al. (CMS Collaboration), Phys. Lett. B 718, 795 (2013).

[16] B. Abelev et al. (ALICE Collaboration), Phys. Lett. B 719, 29 (2013).

[17] G. Aad et al., (ATLAS Collaboration), arXiv: 1212.5198

[18] U. W. Heinz, Concepts of heavy ion physics, CERN-2004-001-D, 2004.